\documentclass[runningheads]{svmult}

\usepackage{makeidx}   

\usepackage{graphicx}  

\usepackage{subeqnar}  

\usepackage{multicol}  

\usepackage{physprbb}  

\makeindex             

\def \deg          {\hbox{$^{\circ}$}}   

\def \htwo         {\hbox{H$_2$}}
\def \hii	   {\hbox{\sc H$\,$ii}}

\def \kms          {\hbox{km$\,$s$^{-1}$}}
\def\approxlt{\lower.2em\hbox{$\buildrel < \over \sim$}}
\def\approxgt{\lower.2em\hbox{$\buildrel > \over \sim$}}
\def \lco          {\hbox{$L_{\rm CO}$}}
\def \lhcn          {\hbox{$L_{\rm HCN}$}}
\def \lir          {\hbox{$L_{\rm IR}$}}
\def \lfir          {\hbox{$L_{\rm  FIR}$}}
\def \ls           {\hbox{L$_{\odot}$}}
\def \Lsun         {\hbox{L$_{\odot}$}}           

\def \ms           {\hbox{M$_{\odot}$}}           
\def \sm           {\hbox{M$_{\odot}$}}           

\def \mvt          {\hbox{$M_{\mbox{\tiny VT}}$}}
\def \Lsun         {\hbox{L$_{\odot}$}}           
\def \Msun         {\hbox{M$_{\odot}$}}           

\begin{document}

\title*{Extreme Starbursts and Molecular Clouds in Galaxies}
%

\toctitle{Extreme Starbursts}

%

\titlerunning{Extreme Starbursts}

%

\author{P. M. Solomon}

%

\authorrunning{P. M. Solomon}

%

\institute{Physics and Astronomy, SUNY, Stony Brook, NY 11794, USA}

\maketitle              

\begin{abstract}

The extraordinary starbursts found in ultraluminous IR galaxies occur in
molecular gas concentrated in compact very massive ``clouds'' which we call
``Extreme Starbursts''. They have one thousand times the mass but are only a
few times larger than GMCs. High-mass star formation in sufficiently dense and
massive structures does not disrupt further star formation; it is a runaway
process.  Star formation remains embedded in the molecular gas and there is
little or virtually no optical--UV radiation.  In the early universe extreme
starbursts may be more frequent and they may be the mode of star formation in
high redshift submillimeter sources.
\end{abstract}

\index{Extreme Starbursts}

\section{Introduction}

All star formation takes place in molecular clouds and most star formation
takes place in Giant Molecular Clouds (GMCs), the most massive objects in the
Galaxy \cite{SS80} and the dominant form of the molecular interstellar
medium. The distribution of GMCs in the Milky Way molecular ring \cite{SS75},
as traced by CO emission, is very different from that of atomic hydrogen and
similar to that of \hii\ regions, demonstrating that it is not the ISM
distribution as a whole that governs star formation rates but the distribution
of molecular gas. An understanding of star formation rates and starbursts in
galaxies requires an understanding of the physical conditions in GMCs and their
relation to galactic dynamics. Due to the limited resolution of current
millimeter wave arrays the Milky Way is the only galaxy in which a large number
of GMCs have been identified and mapped and we must turn to Milky Way GMCs to
examine star formation efficiencies on the scale of individual clouds.

\index{giant molecular clouds}
\index{GMCs}

The high-mass star formation rate within and on the edge of GMCs can be
estimated from the far infrared luminosity emitted by dust heated by embedded
OB stars. The association of molecular clouds and FIR emission from OB star
formation regions in a section of the Milky Way is shown in Figure 1.
The mass of molecular gas can be determined from the CO luminosity
or, more accurately for individual clouds, from the virial theorem utilizing
the CO kinematics. The ratio of the FIR luminosity to the CO luminosity,
\lfir/\lco, or to the cloud mass, \lfir/\mvt, is a measure of the rate of star
formation per solar mass of the cloud and is an indicator of star formation
efficiency.  The star formation rate is $\dot M_* \sim (2.5 \times
10^{-10})~\lir$ [\ms/yr] and the gas depletion time is $\tau = (M_{cloud}/{\dot
M_*)}$ years \cite{GH86}. A FIR and CO survey of 60 GMCs \cite{MS88} shows that
for active OB star forming clouds \lfir/\lco\ is typically $\sim 15$ and
typical molecular gas depletion times are $\approx 10^{9}$ years.  Even in the
most active individual clouds (M17A and W49), \lfir/\lco\ = 100 and the gas
depletion time of $2 \times 10^8$ years is two orders of magnitude greater than
the cloud dynamical timescale. By this measure star formation is clearly not
very efficient in ordinary GMCs.

\begin{figure}[t]

\begin{center}

\includegraphics[width=.635\textwidth,angle=-90]{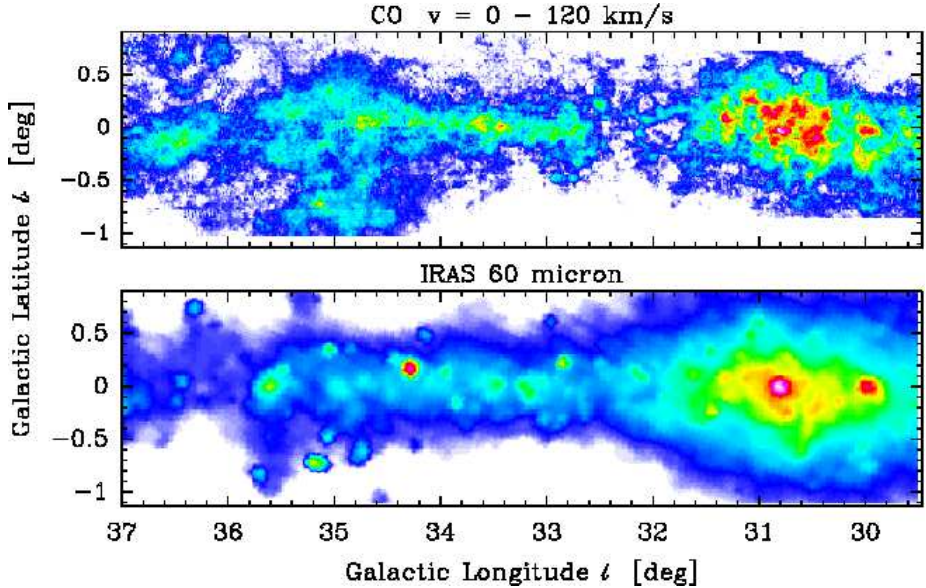}

\end{center}

\caption[] {Molecular clouds and FIR emission from a section of the inner Milky
Way. The CO (1--0) emission in this view includes all velocities along the line
of site blending many clouds. All of the strong FIR emission is associated with
individual clouds and embedded \hii\ regions. The CO is from data obtained at
FCRAO (T. Mooney) with 50 arc second resolution}

\label{eps0}

\end{figure}

\index{starbursts}
\index{infrared}

\begin{figure}

\begin{center}

\includegraphics[width=.90\textwidth]{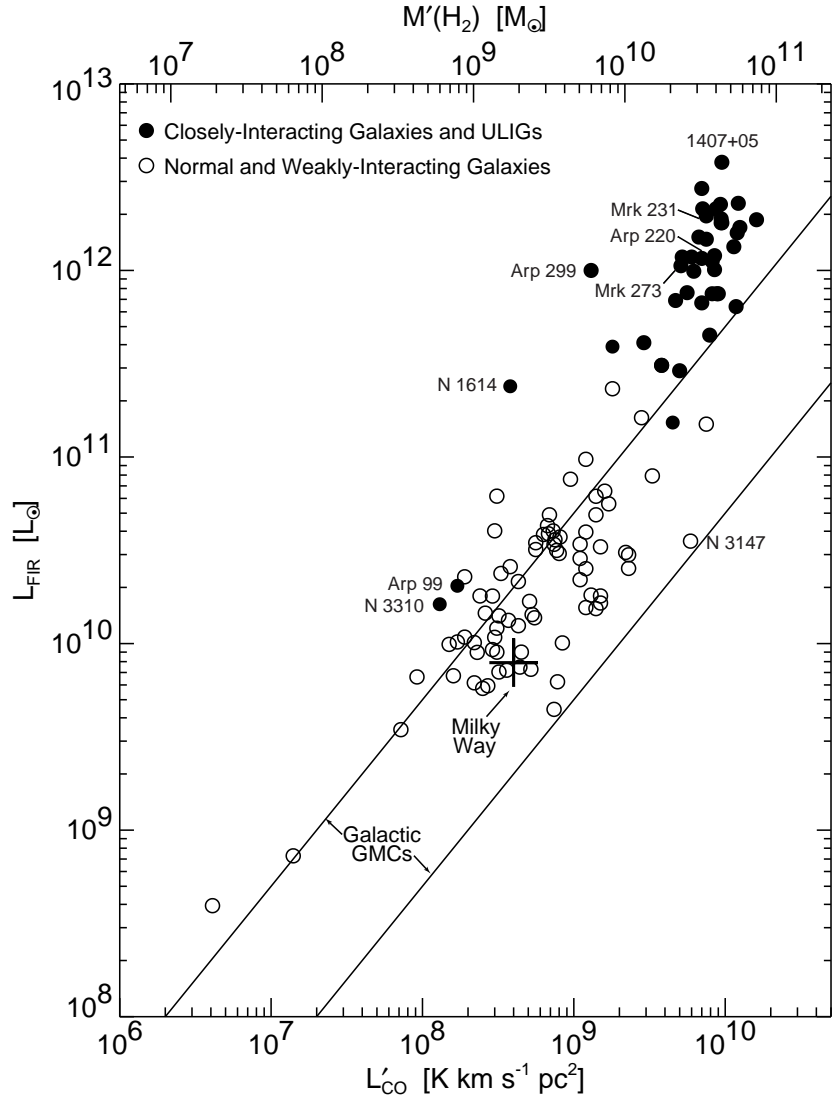}

\end{center}

\caption[]{FIR luminosity vs. CO (1--0) luminosity (lower scale) and molecular
gas mass (top scale). (Adapted from \cite{SS87} and \cite{SDRB97}).
Solid circles indicate closely-interacting galaxies, mergers and
ultraluminous galaxies; open circles are isolated and weakly-interacting
galaxies. The solid lines labeled GMCs bracket the \lfir/\lco\ for Galactic
giant molecular clouds with active OB star formation. The top axis is the
molecular, \htwo, mass assuming a Milky Way CO to \htwo\ conversion
factor. This is an overestimate for ULIGs which have an even higher ratio of
$\lfir/M(\htwo)$. Ultraluminous galaxies have a higher $\lfir/M(\htwo)$ than
any Galactic GMC}

\label{eps1}

\end{figure}

The efficiency of star formation in galaxies must be measured relative to the
potential for star formation determined by the available mass of molecular gas.
Normal spiral galaxies, even those with moderately high IR luminosities, have a
range of \lfir/\lco\ similar to GMCs, indicating star formation efficiencies
similar to those of the Milky Way. In normal spirals, the rate of star
formation is proportional to the mass of molecular gas. In contrast, far IR
starbursts, luminous infrared galaxies (LIGS) and ultraluminous infrared
galaxies (ULIGs) have \lfir/\lco\ ratios 3 to 20 times higher than normal
spirals. Most of these are closely interacting and merging systems
\cite{SAND88}. CO observations of 30 IRAS identified ultraluminous galaxies out
to z = 0.3 show that while all without exception have high CO luminosities,
they also all have abnormally high star formation efficiencies \cite{SDRB97}.
While their CO luminosities are at the very high end of normal galaxies, their
FIR luminosities are more than an order of magnitude above normal spirals. This
situation is summarized in Figure 2, which shows that the star formation
efficiency of ULIGs as indicated by \lfir/\lco\ is not only higher than in
normal spirals, but is higher than that of any individual GMC. CO luminosity
and molecular mass are not very good indicators of high-mass star formation
rates. \lco\ is not actually very well correlated with \lfir. When infrared
luminous galaxies are included, \lfir/\lco\ ranges over a factor of 100. This
large dispersion is further complicated by the interpretation of CO emission
from ULIGs in terms of \htwo\ mass.

\index{star formation efficiency}

\index{molecular hydrogen}
\index{CO observations}

The large nuclear concentration of molecular gas in ultraluminous galaxies has
been mapped in the millimeter lines of CO by several groups during the past
decade. Scaling to the CO signal strengths from Milky Way molecular clouds,
however, soon led to a paradox for many of the sources --- the estimated gas
mass was equal to or larger than the dynamical mass indicated by the linewidths
and source size. For Arp~220, for example, nearly all of the mass in the
central few hundred parsecs was in the form of molecular gas \cite{SC91}. To
resolve this dilemma, we \cite{DSR93,SDRB97} showed that in the extreme
environment in the central few hundred pc of ultraluminous galaxies, much of
the CO luminosity must come from an inter-cloud medium that fills the whole
volume, rather than from clouds bound by self gravity, and therefore the CO
luminosity traces the geometric mean of the gas mass and the dynamical mass,
rather than just the gas mass. This has the effect of lowering the molecular
mass for a given CO luminosity and increasing the star formation efficiency.

What are the physical differences between the molecular gas in ULIGs and normal
spirals which account for the extraordinary infrared starbursts? To answer this
question, we have carried out comprehensive observations of the global
properties of the molecular gas in ULIGs and also of the detailed morphology
and kinematics.

\section {Dense Molecular Gas: HCN, a Molecular Starburst Indicator}

Until recently, CO has been the only molecular tracer systematically observed
in a wide sample of galaxies, particularly ULIGs. CO traces molecular hydrogen,
\htwo\ at densities $\geq 300$~cm$^{-3}$, typical of densities in GMCs, but far
below the densities in GMC cloud cores, the actual sites of star formation. To
measure the mass of \emph{dense} molecular gas it is essential to use a
molecular transition with a much higher density threshold. A particularly
useful molecule in this respect is HCN, which has a moderate abundance but a
high dipole moment, and therefore a short lifetime and requires a large \htwo\
density n(\htwo) $\geq 3 \times 10^4$ cm$^{-3}$ for significant excitation and
emission. Initial measurements of the HCN (1--0) line from 5 ULIGs \cite{SDR92}
showed extraordinarily strong emission. Arp~220, Arp~193, Mrk~231, and NGC~6240
all showed HCN line luminosities greater than the CO luminosity of the Milky
Way, and about 30 times higher than the HCN luminosity of normal spirals. The
unexpected huge HCN luminosity and the high ratio of \lco/\lhcn\ indicate that
a large fraction of the total molecular gas in ULIGs is at a high density
similar to that in star forming cloud cores, rather than the envelopes, of
GMCs.

\index{HCN}
\begin{figure}[b]

\begin{center}

\includegraphics[width=.55\textwidth,angle=90]{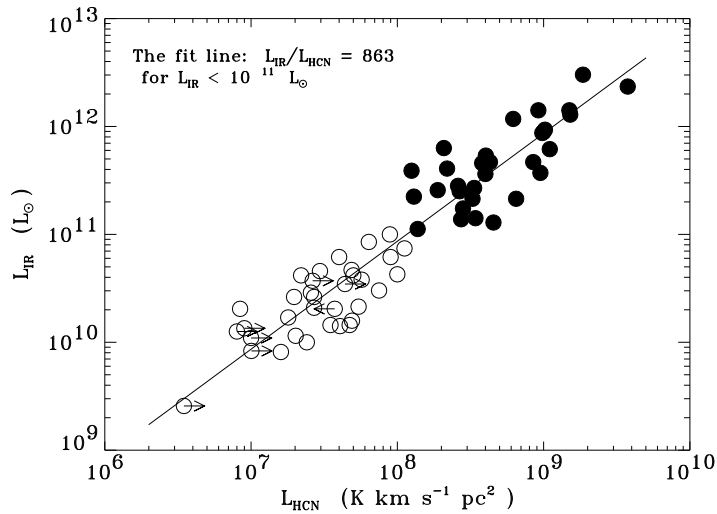}

\end{center}

\caption[]{FIR luminosity vs. HCN (1--0) line luminosity for 65 galaxies (from
\cite{G96} and \cite{GS2001}). Solid circles represent luminous and
ultraluminous galaxies with $\lfir \geq 10^{11}$ \ls. Open circles are all
galaxies with $\lfir < 10^{11}$ \ls. Ultraluminous and normal
galaxies fall along the same line with a constant ratio of \lfir/\lhcn.  HCN
emission traces dense molecular gas with n(\htwo) $\geq 3 \times 10^4$
cm$^{-3}$}

\label{eps2}

\end{figure}

Although the HCN sample contained only 10 galaxies, the results indicated that
the ratio of far infrared to HCN luminosity is similar in ultraluminous
galaxies and normal spirals, including the Milky Way, which suggests that the
star formation rate per solar mass of \emph{dense} gas is independent of the
infrared luminosity. Subsequent HCN measurements (Figure 3) of the global HCN
luminosity in 60 galaxies covering 3 orders of magnitude in IR luminosity
\cite{G96,GS2001} confirm this result, demonstrating that the \lfir\ --
\lhcn\ correlation is substantially better than that for \lfir\ -- \lco, and
show that \lfir/\lhcn\ is almost independent of FIR luminosity (There is a very
small effect amounting to a factor of 1.7 for the most luminous galaxies). This
indicates that the star formation rate is proportional to the dense molecular
gas content of a galaxy, and is also evidence that the power source in ULIGs is
primarily star formation, not AGNs. Normal spiral galaxies clearly powered by
star formation have the same ratio of far infrared luminosity to HCN luminosity
as ultraluminous galaxies. The dense molecular gas in normal spirals and
ultraluminous galaxies is the star forming material being processed into stars
with equal efficiency in all galaxies. HCN luminosity is a star formation
indicator.
\index{Ultraluminous Galaxies}
\index{ULIGs}
\section {Rotating Nuclear Rings and Extreme Starbursts \protect\newline in
Ultraluminous Infrared Galaxies}

High spatial resolution CO maps of ULIGs (eg \cite{SC91,YS95,DS98}) 
all show evidence of central molecular concentrations and rotation. A
survey with the IRAM Array of the kinematics of 10 ULIGs included 5 galaxies
mapped at the unprecedented resolution of 0.5 arc seconds \cite{DS98}.
Analysis of the CO intensity distribution and kinematics utilizing models of CO
excitation and radiative transfer yields a detailed kinematic picture dominated
by rotating rings or discs on a scale of a few hundred parsecs (eg Mrk~231,
Mrk~273, Arp~220, Arp~193) up to 2 kpc (for VII~Zw31), with compact
\emph{Extreme Starburst Regions} embedded in the rings or disks. An example of
the CO (2--1) data is shown in Figure 4 for Arp~220, the nearest ULIG, and
therefore the object with the highest spatial detail. Most of the CO emission
originates from a disk rotating at 330 km/s with a characteristic radius of 340
pc (1 arc second); the strongest CO emission and most of the 1.3 mm dust
continuum emission, indicative of the luminosity source, originates from the
two smaller sources Arp~220 East and West coincident with the extended
nonthermal radio continuum peaks, usually interpreted as the ``nuclei'' of the
merging galaxies. Each of these objects shows its own internal kinematics; the
East source shares a kinematic axis with the main disk and the West source has
a kinematic axis rotated by 110$\deg$ from the East source and main disk
axis. (see Figure 21 in \cite{DS98}). A map with similar resolution obtained at
OVRO \cite{SAK99} yields similar results, but with the East and West kinematic
axes offset by 180$\deg$ interpreted as counter rotating nuclei. A full
interpretation of the relation between these two sources and the larger disk
will have to wait for even higher resolution data.

\begin{figure}

\begin{center}

\includegraphics[width=.95\textwidth,angle=180]{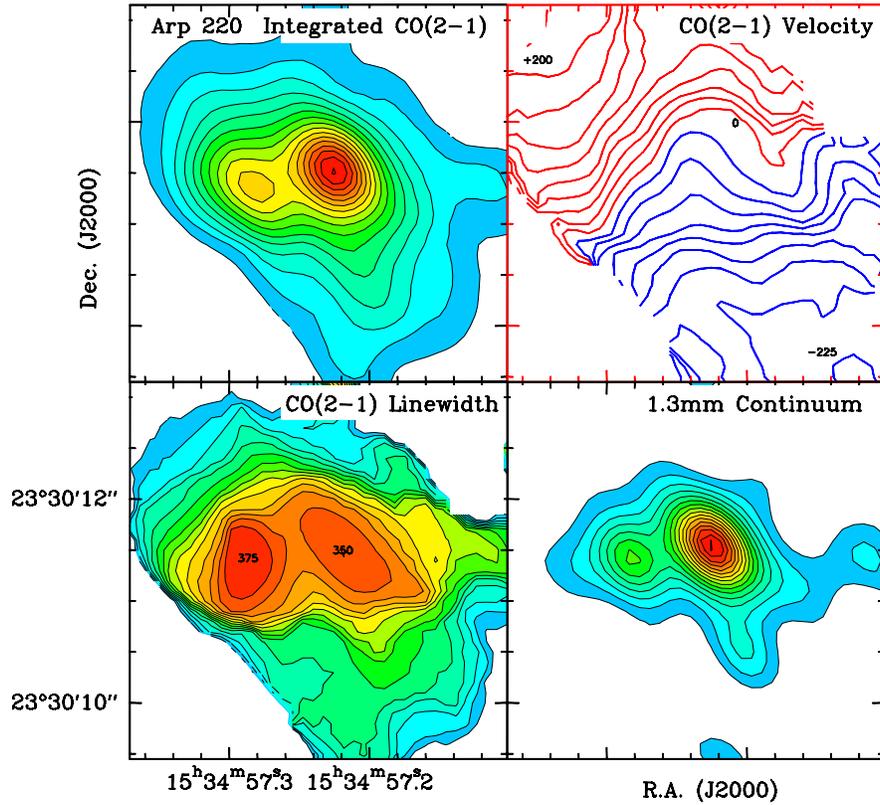}

\end{center}

\caption[]{{\bf Arp 220 :} CO (2--1) integrated intensity, velocity, linewidth
(FWHM), and the 1.3\,mm continuum.  CO integration limits: ($-320,
+300$\,\kms).  Beam $= 0''.7\times 0''.5$.  One arc second corresponds to 350
pc. The two sources of extreme starbursts can be seen in the integrated CO
intensity and the 1.3 mm dust continuum where they completely dominate the
emission.The strong deviations from circular rotation in the disk, indicated by
the sharp twists in the velocity contours, are due to the West source which is
rotating at an angle shifted by 110$\deg$ from the main disk (from
\cite{DS98})}

\label{eps3}

\end{figure}
\index{Arp220}
An analysis of the molecular mass, total dynamical mass and mass of young stars
required to account for the extreme starburst in the East and West sources
\cite{DS98} shows that most of the mass must be in molecular gas and young
stars, not in old stars from a pre-existing nuclear bulge. The characteristic
radii of these extreme starburst regions are only 70 and 110 pc, with total
dynamical masses determined from a virial theorem analysis of rotation of
about 3 and 1.5 $\times 10^9$ \sm, yet each of these regions is producing
between 3 and 5 $\times 10^{11}$ \ls.
\index{Mrk273}
Two other Extreme Starburst Regions have been identified in Mrk~273 and
Arp~193.  The high-resolution CO (2--1) maps show a remarkable molecular-line
source in the Mrk~273 nuclear disk --- a bright, $0.35''\times <0.2''$ CO core,
that resembles the West nucleus of Arp~220.  This is the most luminous extreme
starburst region identified in the sample of 10 galaxies. It has an infrared
luminosity of about $6 \times 10^{11}$\,\Lsun, generated from a current
molecular mass of $1 \times 10^9$\,\sm\ in a region with a radius of only 120
pc (see Table 1).  The extended nonthermal continuum emission \cite{CON91}
coincident with a high mass of dust and gas leaves little doubt that this
region is powered by star formation.  To put this in perspective, the entire
molecular core has a radius about 5 times that of a very IR luminous Milky Way
GMC (for example W51), but with about 3,000 times the molecular mass and
$\approx \, 10^5$ times the IR luminosity from OB stars.

\begin{table}[t]
\caption{Properties of Extreme Starburst Regions}
\begin{center}
\setlength\tabcolsep{10pt}
\begin{tabular}{lcccc}
\hline\noalign{\smallskip}
\null&Arp\,193&Mrk\,273&Arp\,220&Arp\,220\\
\null&SE core&core&west&east\\
\noalign{\smallskip}
\hline
\noalign{\smallskip}
Reference radius: & & & &\\
R(pc)             & 150 & 120 & 68 & 110 \\
Gas mass:         & & & & \\
$M_{\rm gas}(<R)$ (10$^9$\,\Msun ) &0.6    &1.0      &0.6	&1.1 \\
Mean gas density: & & & &\\
$<N({\rm H}_2)>$ (cm$^{-3}$)    	
&$2\times 10^3$    &$5\times 10^3$  	&$2\times 10^4$	&$8\times 10^3$	\\
Total mass: & & & &\\
 $M_{\rm tot}(<R)$ (10$^9$\,\Msun )	 	
					&1.4 	&2.6	&1.4 	&3.2 	\\
Estimated mass in new stars: & & & & \\
$M_{\rm new\star}(<R)$ (10$^9$\,\Msun )	    
					&0.8    &1.6    &0.8	&2.1	\\
Luminosity: & & & & \\
$L_{\rm FIR}(<R)$ (10$^{12}$\,\Lsun )       	
			&0.2    &0.6    &0.3-0.5    &0.2-0.4	\\
Luminosity to mass ratio: & & & & \\
$L_{\rm FIR}/M_{\rm new\star}$ (\Lsun / \Msun)  
					&300	&360    &380	&100	\\
\noalign{\smallskip}
\hline
\noalign{\smallskip}
\end{tabular}
\end{center}
\label{Table1}
\end{table}
\index{30 Doradus}

Table 1 (adapted from Table 12 of \cite{DS98}) lists the
properties of 4 Extreme Starburst Regions identified in the 3 closest galaxies
in the sample. They are the most prodigious star formation events in the local
universe, each representing about 100 times as many OB stars as 30 Doradus. We
are observing these objects near the peak of their star formation, when about
half of the gas has been turned into stars. The duration of the starburst,
limited by the molecular gas supply and the current star formation rate is
about 5 to 10 $\times 10^6$\,yrs. This short lifetime is consistent with the
even shorter dynamical timescale of these compact ($\approx $\, 100 pc)
starbursts. These are not only extreme starbursts, they are extremely efficient
starbursts. The high L/M for these starbursts also suggests that they may
require a high-mass IMF.

The high average density in the Extreme Starburst Regions solves the puzzle of
the origin of the extraordinary HCN emission associated with ultraluminous
galaxies discussed above.  In Arp~220, the HCN lines have the same velocities
as the East and West ``nuclei''. The East and West ``extreme starbursts'' alone
account for most of the HCN emission.  These dense, compact sources have a
hydrogen column density of $0.6 \times 10^{25}$\,cm$^{-2}$ and mean density of
20,000 cm$^{-3}$, enough to thermalize the lower rotational levels of HCN by a
combination of collisions and radiative trapping even if the average density is
the local density.  Within the Arp~220 East and West sources, the HCN emission
may thus have the same intrinsic brightness temperature as the CO (2--1)
emission. Using the observed sizes and linewidths, the HCN luminosity of these
two regions alone is $\lhcn \approx 7 \times 10^8 $ K\,\kms\,pc$^2$, which
is 3/4 of the observed total \cite{SDR92}.  These two regions thus emit only
1/4 of the CO luminosity but most of the HCN luminosity.  It is likely that
high density, extreme starburst regions exist in almost all ultraluminous
galaxies and are the source of most of the star formation and most of the HCN
emission.  The origin of the HCN emission in the high density gas of the
Extreme Starburst Regions directly relates the HCN emission to star formation.

\section{Summary}

The extraordinary starbursts found in ultraluminous IR galaxies occur in
molecular gas concentrated in compact very massive ``clouds '' which we call
``Extreme Starbursts''. They have one thousand times the mass of GMCs, but are
only a few times larger. The entire structure, containing about a billion solar
masses of molecular gas, has an average \htwo\ density characteristic of
molecular cloud cores which represent only a few percent of the molecular mass
in ordinary GMCs. It appears that high-mass star formation in sufficiently
dense and massive structures does not disrupt further star formation. Since the
star formation remains embedded in the molecular gas, there is little or
virtually no optical--UV radiation escaping, and extreme star formation can be
traced only in the far infrared.  If the density and mass are sufficiently
great, star formation is a runaway process. In the local universe this occurs
primarily in galaxy mergers. In the early universe extreme starbursts may be
more frequent and they may be the mode of star formation in the population of
high redshift submillimeter sources.






%






\end{document}